\documentstyle[aps,prc]{revtex}
\draft

\newcommand{\ba}{\begin{eqnarray}}
\newcommand{\ea}{\end{eqnarray}}
\input{psfig}

\begin{document}
\pagestyle{plain}

\title{On the dominance of $J^P=0^+$ ground states
in even-even nuclei from random two-body interactions}

\author{R. Bijker$^1$, A. Frank$^{1,2}$, and S. Pittel$^3$}

\address{$^{1}$Instituto de Ciencias Nucleares, 
Universidad Nacional Aut\'onoma de M\'exico, 
Apartado Postal 70-543, 04510 M\'exico, D.F., M\'exico\\
$^{2}$Instituto de F{\'{\i}}sica, Laboratorio de Cuernavaca,  
Apartado Postal 139-B, 
Cuernavaca, Morelos, M\'exico\\
$^{3}$ Bartol Research Institute, 
University of Delaware, Newark, DE 19716, U.S.A.}

\date{April 28, 1999}

\maketitle

\begin{abstract}

Recent calculations using random two-body interactions showed a 
preponderance of $J^P=0^+$ ground states, despite the fact that 
there is no strong pairing character in the force. We carry out 
an analysis of a system of identical particles occupying orbits 
with $j=1/2$, 3/2 and 5/2 and discuss some general features of the 
spectra derived from random two-body interactions. We show that for 
random two-body interactions that are not time-reversal invariant 
the dominance of $0^+$ states in this case is more pronounced, 
indicating that time-reversal invariance cannot be the origin of 
the $0^+$ dominance.  

\end{abstract}

\begin{center}

PACS number(s): 05.30.-d, 05.45.+b, 21.60.Cs, 24.60.Lz

\end{center}

\section{Introduction}

In a recent paper\cite{JBD}, Johnson {\em et al} discussed the 
low-energy spectra of many-body even-even nuclear systems arising 
from random two-body interactions. Surprisingly, their results showed 
a preponderance of $J^P=0^+$ ground states for these nuclei, despite 
the fact that there is no obvious pairing character in the assumed 
random forces. This is contrary to the traditional folklore whereby 
the favoring of $0^+$ ground states is thought to be a reflection of 
nuclear pairing arising from the short-range nuclear force. 

What is it that produces this preponderance of $0^+$ ground states 
in even-even many-body systems? In this work, we carry out a detailed 
analysis of a system of $N=2$, 4 and 6 identical particles occupying 
orbits with $j=1/2$, 3/2 and 5/2~. We address one of the oft-stated 
suggestions, that it may arise because of the time-reversal invariance 
of the random Hamiltonian. Since time-reversed states play an 
important role in the formation of correlated $0^+$ (Cooper) pairs 
which in turn can give rise to favored collective many-body states, 
it is conceivable that time-reversal invariance may contain 
a built-in preference for $J^P=0^+$ many-body ground states.

\section{Random two-body interactions}

To see whether this is indeed the case, we have carried out an 
analysis very similar to that of \cite{JBD}, but now relaxing the 
assumption of time-reversal invariance in the random two-body 
interaction. This can be done by introducing Gaussian Unitary 
Ensembles (GUE's) rather than Gaussian Orthogonal Ensembles (GOE's) 
to randomly generate the two-body matrix elements.

More specifically, to investigate the effect of time-reversal symmetry 
breaking we consider a two-body Hamiltonian matrix of the form 
\cite{Brody,French} 
\ba
H_{\alpha\alpha'} &=& \frac{S_{\alpha\alpha'}+
i \epsilon A_{\alpha\alpha'}}{\sqrt{1+\epsilon^2}} ~,
\ea
where $S$ and $A$ are real symmetric and antisymmetric random matrices, 
respectively, and $\alpha$ and $\alpha'$ label the two-body states 
$| (j_1 j_2)JT \rangle$ with angular momentum $J$ and isospin $T$.  
The matrix elements $S_{\alpha\alpha'}$ and $A_{\alpha\alpha'}$ are 
chosen independently using a Gaussian distribution of random numbers 
with zero mean and variances 
\ba
\left< S_{\alpha\alpha'}^2 \right> &=& 
v^2_{JT}(1+\delta_{\alpha\alpha'}) ~,
\nonumber\\
\left< 
A_{\alpha\alpha'}^2 \right> &=& v^2_{JT}(1-\delta_{\alpha\alpha'}) ~.
\ea
Here $<>$ denote ensemble averages. For $\epsilon=0$ and 1 they 
correspond to GOE and GUE, respectively. The Hamiltonian is 
time-reversal invariant if the two-body matrix elements are real, 
{\it i.e.} $\epsilon=0$. The breaking of time-reversal symmetry can 
be studied by taking $0 \leq \epsilon \leq 1$. For a given value of 
$J$ and $T$, the above ensemble for two-body interactions gives a 
semicircle level density. The normalization was chosen such that the 
radius of this semicircle distribution does not depend on $\epsilon$ 
\cite{French}.  

The variances $v^2_{JT}$ depend on the particular ensemble chosen 
in the calculations. For the Two-Body Random Ensemble (TBRE) of 
\cite{FW} the variances are independent of the angular momentum $J$ 
and isospin $T$ 
\ba
TBRE &:& \hspace{1cm} v^2_{JT} \;=\; {\bar{v}}^2 ~, 
\label{tbre}
\ea
whereas for the Random Quasiparticle Ensemble (RQE) of \cite{JBD},  
which is obtained by requiring that the ensemble be invariant under 
particle-hole conjugation, the higher values of $J$ and $T$ are 
suppressed with respect to the lower ones 
\ba
RQE &:& \hspace{1cm} 
v^2_{JT} \;=\; \frac{{\bar{v}}^2}{(2J+1)(2T+1)} ~. 
\label{rqe}
\ea
Other choices for the ensembles of two-body matrix elements are 
discussed in \cite{Johnson}. In this paper we study the ensembles 
RQE and TBRE. 

\section{Results}

As a model space we take that of $N$ identical particles in the $sd$ 
shell, which consists of single-particle orbitals with $j=1/2$, 3/2 
and 5/2~. The case of $N=6$ particles is one of the examples 
considered in \cite{JBD,Johnson} and referred to as corresponding to 
the nucleus $^{22}$O which has 6 active neutrons in the $sd$ shell. 
For identical particles the isospin is the same for all states, 
and hence does not play a role. 

\subsection{RQE vs TBRE}

Before turning to the issue of time-reversal invariance and its 
effect on the fraction of $0^+$ ground states, we first consider some 
general issues regarding GOE ensembles. In these calculations we take 
$\epsilon=0$. In Table~\ref{oxygen} we show the percentage of the 
total number of runs for which the ground state has a given angular 
momentum $J$. For $N=2$ particles and a given value of $J$ the 
ensemble is a GOE, whose level distribution is given by a semicircle 
with radius $R=\sqrt{4d_{J}v^2_{J}}$ (here $d_{J}$ is the dimension). 
The differences between TBRE and RQE arise from the $J$ dependence 
of the variances, see Eqs.~(\ref{tbre}) and (\ref{rqe}). Whereas the 
results for TBRE depend solely on the dimension of the matrices, for 
RQE there is an additional suppression of the higher values of the 
angular momentum by the $J$ dependence of the variances. For $N=2$ 
particles the RQE gives already a $J^P=0^+$ ground state in 64.0 $\%$ 
of the cases, compared to 15.9 $\%$ for TBRE and 21.4 $\%$ of $0^+$ 
states in the model space.  

For $N>2$ particles the ensemble is the so-called embedded GOE, 
in which the $N$-body matrix elements are expressed in terms of 
the two-body matrix elements by the usual reduction formulae.  
Subsequently the two-body matrix elements are chosen randomly 
using either RQE or TBRE. For $N=4$ particles there is a dramatic 
increase in the percentage of $J^P=0^+$ ground states for TBRE to 
55.9 $\%$, whereas the percentage of $0^+$ states in the model space 
is now only 11.1 $\%$. For RQE there is only a slight increase. The 
same holds for $N=6$ particles. For the latter case we confirm the 
results already obtained by Johnson \cite{JBD,Johnson}. 

\subsection{Level distributions} 

\begin{figure}[tbp]
\centerline{\hbox{
\psfig{figure=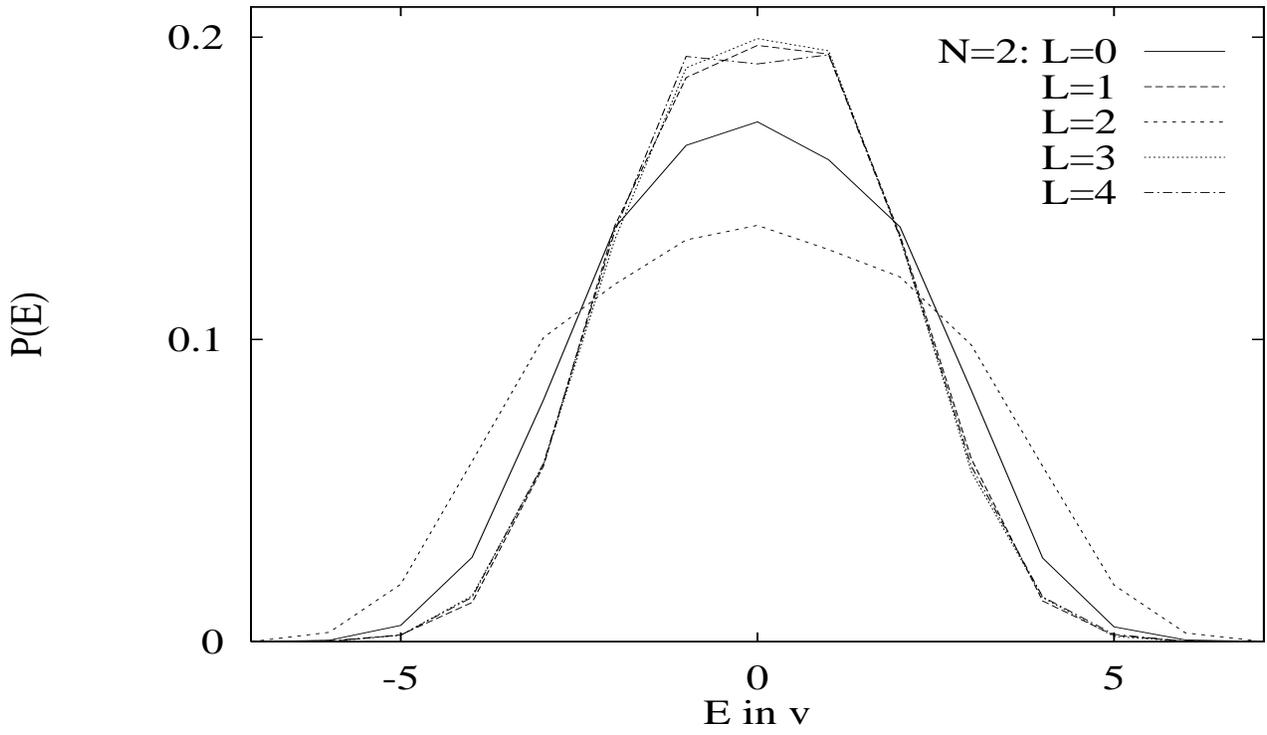,height=0.55\textwidth,width=1.0\textwidth} }}
\vspace{15pt}
\caption[]{Level distributions for $N=2$ particles ($^{18}$O).}
\label{sd2}
\end{figure}

\begin{figure}[tbp]
\centerline{\hbox{
\psfig{figure=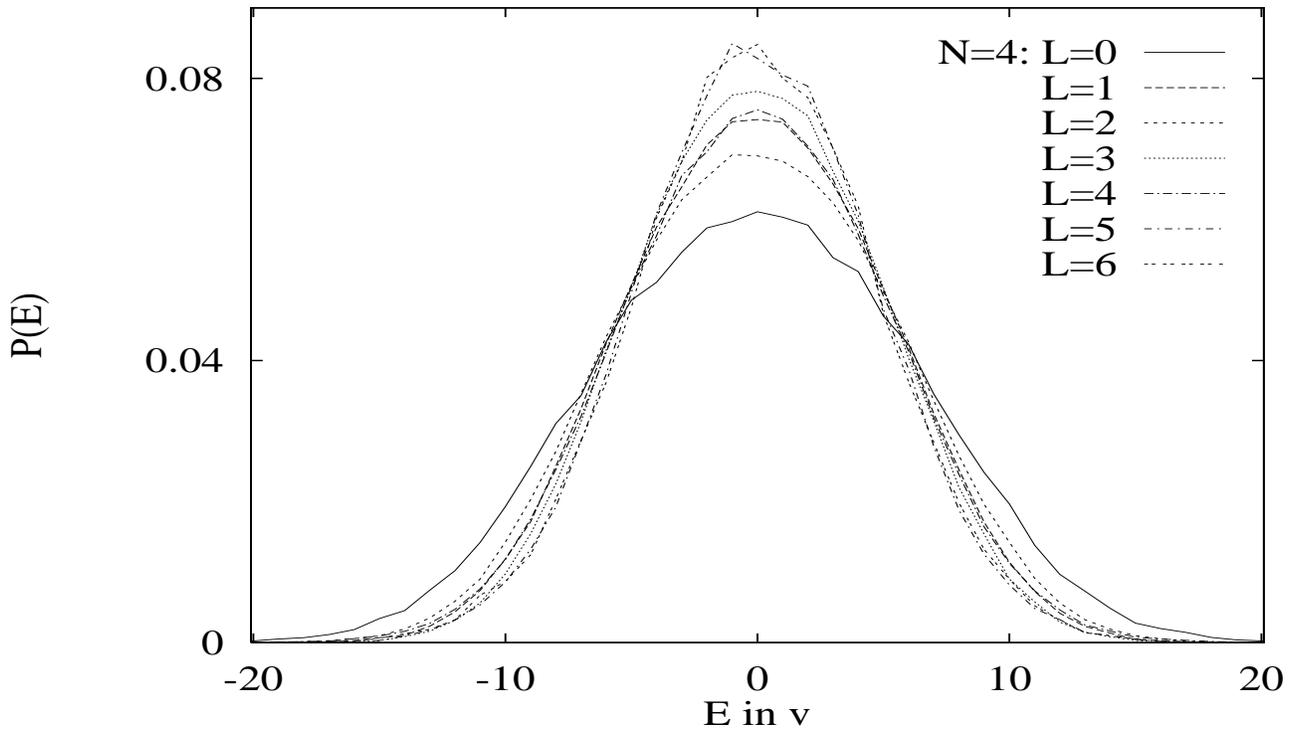,height=0.55\textwidth,width=1.0\textwidth} }}
\vspace{15pt}
\caption[]{Level distributions for $N=4$ particles ($^{20}$O).}
\label{sd4}
\end{figure}

\begin{figure}[tbp]
\centerline{\hbox{
\psfig{figure=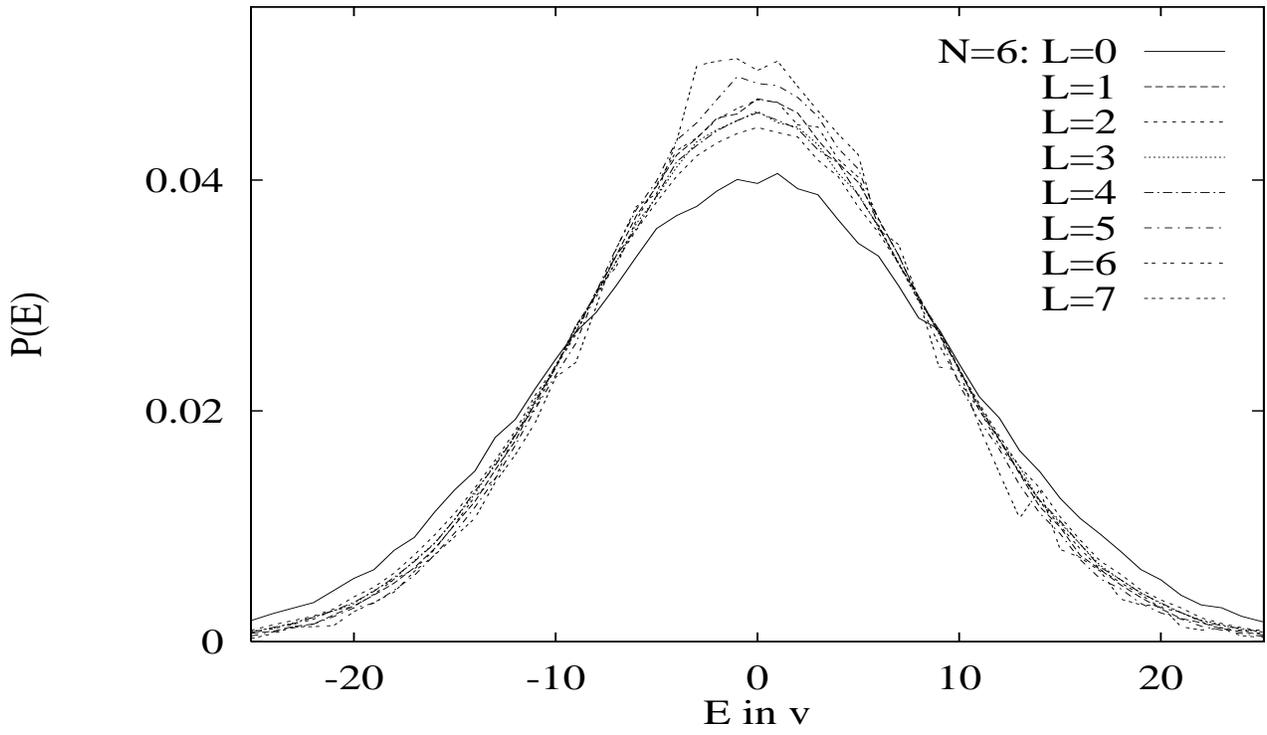,height=0.55\textwidth,width=1.0\textwidth} }}
\vspace{15pt}
\caption[]{Level distributions for $N=6$ particles ($^{22}$O).}
\label{sd6}
\end{figure}

The results of Table~\ref{oxygen} can be better understood by 
examining the energy distributions for each value of the angular 
momentum. Especially interesting are the results for TBRE which 
show a large change between $N=2$, 4 and 6 particles. This is 
illustrated in Figures~\ref{sd2}-\ref{sd6} in which we present 
the corresponding level distributions. The width of the distributions 
can be obtained from 
\ba
w_J &=& \sqrt{\frac{\left< \mbox{Tr}(H^2) \right> 
-  \left< \mbox{Tr}(H) 
\right>^2}{d_J}} ~.
\ea
The values of the widths are given in Table~\ref{widths}. For $N=2$ 
particles the semicircular level distributions have a width 
$w_J=\sqrt{(d_J+1){\bar{v}}^2}$ which only depends on the dimension 
\cite{Brody}. Therefore, $J^P=2^+$ has the largest value of the 
width, followed by $0^+$ and then $1^+$, $3^+$, $4^+$. For $N=4$ and 
6 the level distributions are Gaussian. Here the $J^P=0^+$ has the 
largest width, followed by $2^+$ and the other values of $J^+$. 
This is the result of a competition between the dimensions of the 
Hamiltonian matrices and the correlations in the many-body matrix 
elements arising from the two-body matrix elements. This is another 
manifestation of the dominance of $0^+$ ground states that was 
discussed above. 
 
\subsection{Time-reversal invariance} 

Next we turn our attention to the main point of the present study: 
the breaking of time-reversal invariance. The results of our 
calculations for $N=6$ particles are presented in Table~\ref{O22}. 
For $\epsilon=0$ the Hamiltonian is time-reversal invariant and we 
confirm the results of \cite{JBD,Johnson}. For $0 < \epsilon \leq 1$ 
the time-reversal invariance is broken. We see that the dominance of 
$0^+$ ground states {\em increases} with $\epsilon$ for both RQE and 
TRBE. The increase is from 74.2 to 85.7 $\%$ for RQE and from 
67.7 to 76.8 $\%$ for TRBE. On the basis of these results, we 
conclude that time-reversal invariance of the two-body interaction 
terms is not the origin of the preponderance of $0^+$ ground states 
that emerged in the analyses of \cite{JBD,Johnson}. 

\section{Summary and conclusions}

In this paper, we have investigated the recent observation that 
the spectra of many-body systems with random two-body interactions  
show a predominance of $J^P=0^+$ ground states \cite{JBD}. This 
result is quite robust, and does not depend sensitively on the 
choice of the random matrix ensemble \cite{Johnson}. However, we 
have shown that for $N=2$ particles there is a large difference 
between the Random Quasiparticle Ensemble and the Two-Body Random 
Ensemble. This can be understood from the suppression of the high 
angular momenta in the variances of the two-body matrix elements of 
the RQE, compared to TBRE. For $N=4$ and 6 particles the results for 
the two ensembles become comparable. Since the RQE implies an 
additional suppression of higher angular momenta, we suggest that 
further efforts should concentrate on TBRE Hamiltonians.

A study of the TBRE level distributions for $N=2$, 4 and 6 particles 
in the $sd$ shell shows a rapid change from semicircular 
distributions whose radius depends only on the dimension (for $N=2$) 
to Gaussian distributions (for $N=4$ and 6). In the latter case, the 
widths are determined by a competition between the dimension and the 
many-body dynamics. The widths for all $J \neq 0$ values 
become roughly comparable, whereas the $J=0$ width is larger.

Most importantly, we have shown that the dominance of $0^+$ ground 
states is not a consequence of the time-reversal invariance of the 
two-body force, since the breaking of this symmetry leads to a slight 
increase of the percentage of $0^+$ ground states. It seems that this 
effect arises solely from the differences in the correlations for the 
$n$-body matrix elements for each angular momentum. We are presently 
studying the angular momentum coupling behavior of TBRE Hamiltonians 
in an analytically tractable model \cite{work}. The understanding of 
this problem and more generally that of the spectral properties of 
two-body random ensembles could have significant consequences on 
random matrix analyses of complex many-body systems \cite{Brody}.

\section*{Acknowledgements}

The authors wish to thank Calvin Johnson who brought this problem to
their attention at the XXIInd Oaxtepec Symposium on Nuclear Physics. 
Two of the authors (RB and SP) wish to acknowledge the Institute for 
Nuclear Theory where part of this work was carried out. One of the 
authors (SP) wishes to acknowledge a fruitful discussion with George 
Bertsch during that visit. This work is supported in part by 
DGAPA-UNAM under project IN101997 (RB and AF), and in part by the 
National Science Foundation under grant \# PHY-9600445 (SP).

\clearpage 

\begin{table}[tbp]

\caption[]{Percentage of ground states with angular momentum $J$ for RQE 
and TBRE for $N=2$, 4 and 6 identical particles in the $sd$ shell 
(the 
nuclei $^{18,20,22}$O). 
The results are obtained for 1000 runs with 
$\epsilon=0.0$.} 
\label{oxygen}
\vspace{15pt}
\begin{tabular}{ccrrrr}
& 
& & & & \\
$N$ & $J$ & $d_J$ & basis & RQE & TBRE \\
& & & & & \\
\hline

& & & & & \\
2 & 0 & 3 & 21.4 $\%$ & 64.0 $\%$ & 15.9 $\%$ \\
  & 1 & 2 & 
14.3 $\%$ &  5.3 $\%$ &  4.9 $\%$ \\
  & 2 & 5 & 35.7 $\%$ & 29.3 $\%$ & 
68.3 $\%$ \\
  & 3 & 2 & 14.3 $\%$ &  0.9 $\%$ &  6.1 $\%$ \\
  & 4 & 2 & 
14.3 $\%$ &  0.5 $\%$ &  4.8 $\%$ \\
  & & & & \\
\hline
& & & & & \\
4 & 
0 &  9 & 11.1 $\%$ & 66.6 $\%$ & 55.9 $\%$ \\
  & 1 & 12 & 14.8 $\%$ &  
4.7 $\%$ &  4.9 $\%$ \\
  & 2 & 21 & 25.9 $\%$ & 20.5 $\%$ & 22.7 $\%$ \\
  
& 3 & 15 & 18.5 $\%$ &  0.9 $\%$ &  1.4 $\%$ \\
  & 4 & 15 & 18.5 $\%$ &  
6.6 $\%$ & 12.3 $\%$ \\
  & 5 &  6 &  7.4 $\%$ &  0.5 $\%$ &  1.5 $\%$ \\
  
& 6 &  3 &  3.7 $\%$ &  0.2 $\%$ &  1.3 $\%$ \\
& & & & & \\
\hline
& & & 
& & \\
6 & 0 & 14 &  9.9 $\%$ & 74.2 $\%$ & 67.7 $\%$ \\
  & 1 & 19 & 
13.4 $\%$ &  0.8 $\%$ &  1.3 $\%$ \\
  & 2 & 33 & 23.2 $\%$ & 12.7 $\%$ & 
15.0 $\%$ \\
  & 3 & 29 & 20.4 $\%$ &  5.6 $\%$ &  7.1 $\%$ \\
  & 4 & 26 
& 18.3 $\%$ &  5.7 $\%$ &  6.8 $\%$ \\
  & 5 & 12 &  8.5 $\%$ &  0.4 $\%$ 
&  0.4 $\%$ \\
  & 6 &  8 &  5.6 $\%$ &  0.6 $\%$ &  1.7 $\%$ \\
  & 7 &  
1 &  0.7 $\%$ &  0.0 $\%$ &  0.0 $\%$ \\
& & & & & \\
\end{tabular}

\end{table}

\clearpage
 
\begin{table}[tbp]
\caption[]{Widths of level distributions 
for TBRE.  
The results are obtained for 10000 runs.}
\label{widths}

\vspace{15pt}
\begin{tabular}{cccc}
& & & \\
$J$ & $N=6$ & $N=4$ & $N=2$ 
\\
& & & \\
\hline
& & & \\
0 & 10.16 & 6.24 & 2.00 \\
1 &  8.53 & 5.05 & 
1.73 \\
2 &  9.01 & 5.37 & 2.45 \\
3 &  8.80 & 4.79 & 1.73 \\
4 &  8.80 & 
5.12 & 1.74 \\
5 &  8.27 & 4.65 & \\
6 &  8.61 & 4.69 & \\
7 &  8.07 & & 
\\
& & & \\
\end{tabular}
\end{table}
 
\begin{table}[tbp]

\caption[]{Percentage of $J^P=0^+$ ground states for RQE and TBRE 
for 
$N=6$ identical particles in the $sd$ shell (the nucleus $^{22}$O).  
The 
results are obtained for 1000 runs.}
\label{O22}
\vspace{15pt}

\begin{tabular}{ccc}
& & \\
$\epsilon$ & RQE & TBRE \\
& & \\
\hline
& & 
\\
0.00 & 74.2 $\%$ & 67.7 $\%$ \\
0.25 & 76.3 $\%$ & 69.3 $\%$ \\
0.50 & 
79.7 $\%$ & 71.7 $\%$ \\
0.75 & 83.4 $\%$ & 74.0 $\%$ \\
1.00 & 85.7 $\%$ 
& 76.8 $\%$ \\
& & \\
\end{tabular}
\end{table}

\end{document}